\documentclass[preprint]{aastex}

\usepackage{graphicx}
\usepackage{wrapfig}
\usepackage{natbib}

\newcommand{\be}{\begin{equation}}
\newcommand{\ee}{\end{equation}}
\newcommand{\nn}{\mbox{} \nonumber \\ \mbox{} }
\newcommand{\ba}{\begin{eqnarray}}
\newcommand{\ea}{\end{eqnarray}}

\newcommand{\curl}{{\rm curl\, }}
\newcommand{\A}{{\bf A}}
\newcommand{\E}{{\bf E}}
\newcommand{\B}{{\bf B}}

\renewcommand{\div}{{\rm \,div\,}}

\newcommand\eg{{\it{e.g.\ }}}

\newcommand{\Bf}{{magnetic field}}
\newcommand{\Bfs}{{magnetic fields}}
\newcommand{\Ef}{{electric  field}}
\newcommand{\Efs}{{electric fields}}

\newcommand{\NS}{neutron star}

\newcommand{\EM}{electromagnetic}

\newcommand{\ms}{magnetosphere}
\newcommand{\mss}{magnetospheres}

\newcommand{\LC}{light cylinder}
\newcommand{\Lf}{Lorentz factor}

\begin{document}

\title{Radial diffusion in    corotating \ms\ of  Pulsar PSR J0737-3039B}
\author{Maxim Lyutikov, \\
Department of Physics and Astronomy, Purdue University, 
 525 Northwestern Avenue,
West Lafayette, IN
47907-2036, USA; lyutikov@purdue.edu}
\begin{abstract}
Rich observational phenomenology associated with Pulsar B in 
PSR J0737$-$3039A/B  system resembles in many respects  phenomena observed in  the   Earth and Jupiter \mss, originating  due to the   wind-\ms\ interaction. 
 We consider particle dynamics in the  fast corotating \ms\ of Pulsar B, when the spin period  is shorter  than the third adiabatic period.
 We  demonstrate  that trapped  particles   occasionally experience large radial variations of the L-parameter (effective radial distance) due to the  parametric interaction of the  gyration motion with the large scale  \Efs\ induced  by the deformations of the \ms,  in  what could be called a betatron-induced  diffusion.  The dynamics of particles from the wind of Pulsar A  trapped inside  Pulsar B \ms\  is  governed by Mathieu's equation, so that the   parametrically unstable orbits are  {\it occasionally}  activated; particle  dynamics is not diffusive {\it per se}. 
 The model explains the high  plasma density on the closed field lines of Pulsar B, and the fact that the observed eclipsing region is  several  times smaller than predicted by the hydrodynamic models.
\end{abstract}

\section{Introduction}

The   discovery of  eclipsing binary pulsar system PSR J0737-3039A/B
 \citep{burgay03,lyne}
has been hailed as a milestone both in observational search for an important predicted 
product of stellar evolution, as  an important tool in studying relativistic gravitational physics, and plasma physics in  close environment
of  neutron stars \citep{Breton,2005MNRAS.361.1243P,2006Sci...314...97K,2009CQGra..26g3001K,2008ARA&A..46..541K}. 

In this system, a fast recycled Pulsar A with period $P_A= 22.7$ msec 
orbits
a slower but younger  Pulsar B which has a period $P_B=2.77$ sec in  the tightest binary \NS\ orbit
of $P_b= 2.4$ hours. In addition to testing general relativity, this system provides a truly golden opportunity  to verify and advance our models of pulsars magnetospheres,  mechanisms of generation of radio emission and properties of their relativistic winds. This is  made possible by a fortunate coincidence   that
 the line of sight lies almost
in the orbital plane, with inclination less than half a degree \citep{ransom04,lt05}.

This leads to a number of exceptional observational properties of the system.  The most important one  is that 
 Pulsar A is eclipsed once per orbit, 
for a duration of $\sim 30$ s 
centered around superior conjunction (when Pulsar B is between the observer and Pulsar A).
The width of the eclipse is only a weak function of the observing frequency
\citep{kaspi}. Most  surprisingly, during eclipse the Pulsar A radio flux is
modulated by the rotation of Pulsar B: 
there are  narrow, transparent windows in  which
the flux from Pulsar A rises nearly to the unabsorbed level \citep{mclau04}.
Importantly, the width of the region which causes this periodic modulation is smaller by a factor of $\sim 4-6$ than the estimated size of the magnetosphere
of Pulsar B. 

The basis for understanding  the behavior of the system is provided by the work of 
\cite{lt05} \citep[see also][]{2005MNRAS.362.1078L,2014MNRAS.441..690L}, who constructed a
 model  of A eclipses, which successfully reproduces the eclipse light curves,  down to intricate details. The model assumes that the radio pulses of A are absorbed by relativistic particles populating  the B \ms\ through synchrotron absorption. 
 The modulation of the radio flux during the eclipse is due to 
the fact that -- at some rotational phases of Pulsar B -- the line 
of sight  only passes through open magnetic field lines where absorption
is assumed to be negligible. 
The model explains most of the
properties of the eclipse:  its asymmetric form, the nearly
frequency-independent duration,
and the modulation of the brightness of Pulsar A
at both once and twice the rotation frequency of Pulsar B in different parts
of the eclipse. There are modest deviations between the model and the data near the edges of
the eclipse that  could be used to probe the distortion of magnetic field
lines from a true dipole.

Importantly, the plasma on closed field lines should be very dense, with multiplicity $\sim 10^6$ of the \cite{GJ} density.  
Presence of  high multiplicity,
relativistically hot plasma on the closed field lines of Pulsar B is somewhat surprising, but not unreasonable. 
Dense, relativistically hot plasma can be effectively stored
in the outer magnetosphere, where cyclotron cooling is slow \citep{lt05}.
The gradual loss of particles inward through the cooling radius, occurring on time
scales of millions of Pulsar B periods, can be easily
compensated by a relatively weak  upward flux driven by a fluctuating component
of the current. For example, if suspended material is resupplied at a rate of 
one Goldreich-Julian density per B period and the particle residence time is million
periods, the equilibrium density will be as high
as $10^6 n_{GJ}$.

\section{Plasma entry from wind of A  into \ms\ of B}

There are  two challenges   that the model of \cite{lt05,2014MNRAS.441..690L} requires to be explained. 
One is that closed field lines are populated with hot dense plasma, exceeding the minimal 
Goldreich-Julian density by a factor $\sim 10^5$.  This runs contrary to the conventional view that closed field lines are dead, populated by a cold plasma with minimum Goldreich-Julian density.

Second, the size of eclipsing region is some 4-6 times smaller than the expected size of the wind-confined  \ms\  of Pulsar B \citep{lt05}. For approximately  dipolar field, $B^2 \propto r^{-6}$, this is a large  difference, if one considers the pressure balance.

The current model addresses both these points: particles from the Pulsar A wind diffuse inward from the magnetopause. As a result plasma density increases inward within the \ms. This explains both the high multiplicity  of absorbing plasma and the smaller eclipse size. 

Three  steps   are involved: first particle should penetrate the closed field lines. Then particles diffuse inwards - this is the topic of the present paper. 
Finally particles precipitate due to synchrotron losses. In this paper we address the second question: how does the particle diffusion proceeds in a corotating \ms\ of Pulsar B.
 
 As for particles' entry into the \ms,  there are two generic ways plasma may enter \ms, through the cusp region (reconnection model \citep{Dungey}) and  through plasma instabilities over the whole  interface \citep[dayside reconnection][]{1961CaJPh..39.1433A,1964P&SS...12..273A,1973SSRv...14..511K,1974ApJS...27..261C,1981JGR....8610049S}. 
 The relative importance of two mechanisms depend on the details of the magnetospheric structure 
\citep[see, eg a comparison of Jovian and Earth auroras][]{2021SciA....7.1204Z}. 

The orbital dependence of the pulsed X-ray emission from Pulsar B \citep{2016ApJ...824...87I}  is due to reconnection-induced, orbital phase-dependent penetration of the wind plasma onto closed field lines. The plasma entry through the cusp  depends on the average angle between the cusp normal and the wind \citep{2004spip.book.....K}; in the Double Pulsar system this average angle depends on the orbital location.

Even without cusp penetration,   
observations of the planetary \mss\  indicate that the
 magnetospheric boundary is porous, allowing $\sim 10\%$ of the incoming plasma to penetrate \citep{1971JGR....76..883H}.

One of the key observational and theoretical problems  in planetary  \mss\ is radial diffusion of trapped plasma particles  and the associated formation of  Van Allen  radiation belts \citep{1974ApJS...27..261C,Lanzerotti,2010GeoRL..3722107T,1998JGR...10320487S,2019JGRA..124.8319L}. 
Radial diffusion in the Earth \ms\ requires violation of the third adiabatic invariant associated with particle's drift around the Earth. Typically the third invariant is  violated due to  resonant interaction of the electron drift motion with ultra-low frequency  (ULF, mHetz range)  \EM\ fluctuations. \citep[ULF waves are global oscillations of the earth \Bf, see \eg][]{2005JGRA..11010202U}.

In this paper we consider the question of plasma dynamics in the \ms\ of Pulsar B: ``How does radial diffusion proceed in highly co-rotating  Pulsar B's \ms?''
We demonstrate that  radial diffusion in the Pulsar B \ms\ will proceed in a different way if compared with the case of the Earth and Jupiter: in a  highly corotating regime. Most importantly  {\it the spin frequency  in this case is much larger than the third adiabatic frequency}. This is a new, unexplored regime of radial diffusion.
As we demonstrate in this paper, periodic perturbations of the magnetosphere due to the rotation of Pulsar B ``rotationally pump-in'' particles from larger radii towards the star  without a need for additional scattering by fluctuating disturbances.



\section{Dungey-type  model of corotating \ms}

\subsection{Oscillating Dungey-type \mss}

We start with a mathematically simple  Dungey-type \mss, and extend it to more realistic geometries in \S  \ref{distorted3}. 

The structure of the planetary \mss\ distorted by the Solar wind is a mathematically complicated problem \citep[\eg][]{2002JGRA..107.1179T,2002JGRA..107.1309J}.
\cite{Dungey} constructed a simple, but highly illuminating model of the interaction of the planetary \mss\ with the solar wind. The model approximates the \ms\ as spherical, and provides analytical estimates of the overall structure, like the location of the magnetopause. 
Dungey model of \ms, when the total \Bf\ is represented as a linear sum of dipole fields plus wind's \Bf, is a very successful model of plasma  entry into Earth \ms.  

Importantly, in the case of Pulsar B the confined  \ms\  is in corotation. This makes it similar to the Jovian \ms, where corotation dominates up-to magnetopause, opposite to Earth case where corotation is observed only at low plasma sphere
\citep{1970Icar...13..173B}.

Below we adopt the simple prescription of \cite{Dungey}  for the structure of corotating \ms\ of Pulsar B. 
 Since \Bf\ lines are compresses at the head part of  the \ms, and stretched out in the tail part, we approximate this periodic expansion-contraction as  {\it an  oscillating Dungey \mss}.

This simplification neglects day/night and  dawn-dusk asymmetry within the \mss. Dawn-dusk asymmetry means that the magnetopause on the dawn side is further away than on dusk. This is seen     in simulations  Jovian \ms\ \citep{1998JGR...103..225O}, as well as  in  simulations of Pulsar B \citep{2004cosp...35.4117S}.

In some respect, the relativistic  magnetically dominated \mss\ of pulsar B is simpler than the Earth/Jupiter \mss.
Since both  the pulsar \mss\  and the winds are highly magnetized we can neglect plasma loading, but still allow plasma to carry currents required by the dynamics - a dynamic force-free approximation.  Effects of gravity are not important.  There is no  rotationally supported magneto-disk, where dipolar field lines are stretched out by centripetal force acting on the trapped plasma, as in Jupiter.  There is also no  Io to supply the plasma internally.  Also, Pulsar B does not have the tail plasma sheet: beyond the Pulsar  B's light cylinder interaction of the pulsars proceeds via wind-wind interaction, not of the wind and static magnetic field. \cite[See interesting early discussion by][]{1975SSRv...17..857K}.

\subsection{Mathematics of oscillating   \mss}

Let's approximate  the \ms\  of B as  spherical,  with oscillating radius $R_0 (t)$. The following are the  solutions  for magnetic potential $\Phi$, magnetic flux function $\Psi$, toroidal component of the vector potential $A_\phi$, the \Bf,  the \Ef, and toroidal currents:
\ba && 
\Phi = - \left(\frac{R_0{}^2}{r^2}+\frac{2 r}{R_0} \right)\cos (\theta )
\nn &&
\Psi =  \left(\frac{R_0{}^2}{r}-\frac{r^2}{R_0}\right) \sin ^2(\theta )
\nn &&
A_\phi =  \left(\frac{R_0{}^2}{r^2}-\frac{r}{R_0}\right) \sin (\theta )
\nn &&
\B = \nabla \Phi  =\curl {\bf A} =   \nabla  \Psi  \times \nabla \phi= 
\left\{\cos (\theta ) \left(\frac{2
   R_0{}^2}{r^3}-\frac{2}{R_0}\right),\frac{\sin (\theta )
   \left(\frac{R_0{}^3}{r^3}+2\right)}{R_0},0\right\}
   \nn &&
   \E =\left\{0,0,-\frac{\sin (\theta ) \left(r^3+2 R_0{}^3\right) \dot{R_0}}{r^2
   R_0{}^2}\right\}
   \nn &&
   E_\phi  = - \partial_t A_\phi
   \nn &&
   \nabla \times  \E = - \partial_t \B
   \nn &&
   J_\phi =  \nabla \times  \B  - \partial_t \E = \frac{\sin (\theta ) \left(R_0 \left(r^3+2 R_0{}^3\right) {\ddot{R}_0} -2
   \left(r^3-R_0{}^3\right) \left(\dot{R}_0\right){}^2\right)}{r^2 R_0{}^3}
   \label{1}
   \ea
   where overall normalization and the factor $4 \pi$ have been absorbed into the definitions.
   
   Equation for field line
   \ba &&
   \frac{\left(1-\tilde{r}^3\right) \sin ^2(\theta )}{\tilde{r}}=\frac{1-L^3}{L}
   \nn &&
   \tilde{r} = \frac{r}{R_0}
   \label{l}
   \ea
   Parameter $0< L <1 $ is the  analogue of  the conventional  magnetic shell parameter  in the theory of  radial diffusion, but defined at each moment with respect to the overall radius at that time.
   
 The   \EM\ drift velocity
   \ba &&
   \beta_{EM} = \frac{\E\times \B}{B^2}=
   \left\{\sin ^2(\theta ) \left(5 r^3 R_0{}^3+2 r^6+2 R_0{}^6\right),\sin (2
   \theta ) \left(r^3 R_0{}^3+r^6-2 R_0{}^6\right),0\right\} 
   \times
   \nn &&
   \frac{2 r \dot{R}_0}{-4 r^3 (3 \cos (2 \theta )+1) R_0{}^4+8 r^6 R_0+(3 \cos (2
   \theta )+5) R_0{}^7}
      \nn &&
|\beta_{EM} |=    \frac{\sqrt{2} r \sin (\theta ) \left(r^3+2 R_0{}^3\right)}{\sqrt{-4 r^3 (3 \cos
   (2 \theta )+1) R_0{}^5+8 r^6 R_0{}^2+(3 \cos (2 \theta )+5) R_0{}^8}} {\dot{R}_0}
   \ea
   Fig. \ref{1}. On the surface $\beta_{EM} =\{ {\dot{R}_0},0,0\}$. 
 
Since  \EM\ velocity is perpendicular to \Bf, this is approximately the rate of change of $L$. Importantly, $\beta_{EM} \approx r/R_0$  - it varies slowly within the \mss.  The \EM-induced diffusion is not suppressed by large \Bf\ closer to the star.

  \begin{figure}[h!]
\includegraphics[width=.99\textwidth]{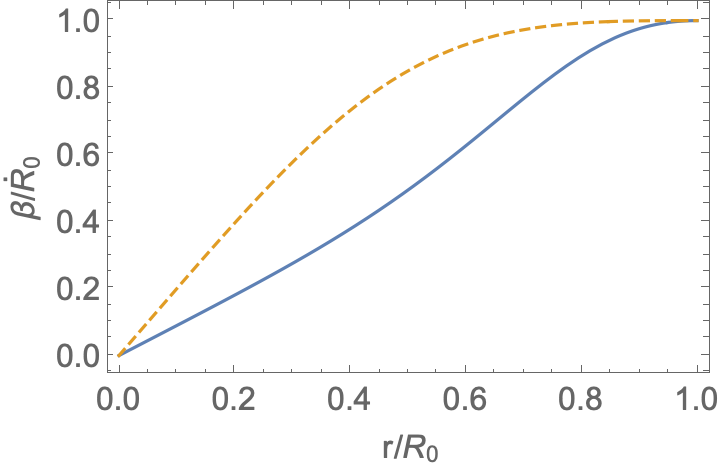}
\caption{Velocity $|\beta_{EM}|$ of  oscillating Dungey's \mss:  top curve for  $\theta = \pi/2$, bottom for $\theta =\pi/4$. This illustrates that the velocity of oscalailateiton is $ \beta \approx r $. }
\label{111} 
  \end {figure}
  
  The above solutions establish the large scale \EM\ field within the \ms.  Next we consider particle dynamics in the given \EM\ fields, first theoretically in \S  \ref{radial},  and then numerically in  \S \ref{Simulations}

\section{Theory of radial diffusion in corotating \mss}
\label{radial} 

\subsection{The three adiabatic invariants }

For particles trapped on closed field lines, there  are  three adiabatic invariants: first associated with cyclotron motion around a field line, second associate with bouncing motion between the magnetic poles, and a third one associated with azimuthal B-cross-grad-B drift. 

In the Double Pulsar, the 
conservations of the first  invariant is assured by  very short cyclotron time even at the  magnetospheric boundary.  
The second adiabatic invariant is mildly conserved in the corotating  \ms, which extends nearly to the \LC.
The bounce period  for relativistic particle \citep[second adiabatic invariant,][Eq. 4.28]{2005igtr.book.....W}
  \be
  \tau_b \sim 5.4 {r \over  c} \leq {2\pi}/\Omega_B
  \ee
  This is somewhat shorter than  the Pulsar B period  at the outer parts of the \ms, and even shorter deeper inside \ms\ --  the second  invariant is mostly conserved,  and is well conserved in the inner regions.


 Thus,  similarly to the case of planetary \mss, radial diffusion in the Pulsar B \ms\ occurs through breakdown of the third adiabatic  invariant: spin-induced variation of the \ms\ lead to changing $L$ parameter.  We consider this process next.
  

   \subsection{Magnetic pumping/betatron induced  diffusion}
   
   As an  illustration  how varying magnetic  field leads to diffusion, let's
   consider  straight \Bf\ changing in time $B_z= B$. (This is a local approximation, hence we are not concerned with the global structure of the fields.) Choosing vector potential in the Landau gauge
   \be
   \A = - \{ B y,0,0\}
   \label{AAA} 
   \ee
   We find the Hamiltonian
   \be
   H = \frac{(p_x +B y)^2 +p_y^2}{2}
   \ee
   The $x$ momentum is a conserved quantity,
   \be
   p_x= p_{x,0}
   \ee
   while equation for $y$-momentum becomes
   \ba &&
   \partial_t p_y =- B(p_{x,0}+ B y)
   \nn && 
   \partial_t  y = p_y
   \ea
   with the  appropriate constants ($c,m_e,e$) set to unity.
   
   For harmonic variation of the field (this is to be used in the expression for vector potential (\ref{AAA})) 
   \be
   B = B_0 + (\delta B) \cos (\Omega t) 
   \ee
   we find
   \ba &&
   \ddot{p}_y ={(\delta B)} \Omega  \sin ( \Omega t) \left(-\frac{2
   \dot{p}_y}{{B_0}+{(\delta B)} \cos ( \Omega t
   )}-{p_{x,0}}\right)-p_y ({B_0}+{(\delta B)} \cos ( \Omega t
   ))^2 \approx 
   \nn &&
   {(\delta B)} \left(-\frac{\Omega  \sin ( \Omega t) \left({B_0} {p_{x,0}}+2
   \dot{p}_y\right)}{{B_0}}-2 {B_0} p_y \cos ( \Omega t)\right)-{B_0}^2 p_y
   \ea
   For small frequencies $\Omega \to 0$ this reduces to Mathieu's equation
   \be
  \ddot{p_y} + {B_0}^2 p_y+2 {B_0} {(\delta B)}  \cos ( \Omega t
   ) p_y =0
   \label{Mathieu}
   \ee
   Mathieu's equation has instability bands \citep{MCLACHLAN}:  {\it  this is the origin of radial diffusion in corotating \mss}.
   
   \subsection{The diffusion coefficient} 
   
   The particle motion within the \ms\ of Pulsar B obeys Mathieu's equation. As parameters of the orbit change, particle motion occasionally becomes unstable, with large changes in the L-parameter. The motion  is stochastic, but non-diffusive, periods of constant L-parameter are intermittent with periods of rapid radial  evolution. To understand time-average behavior we may still use a concept of diffusion (time-averaging  here means that motion is averaged  over sufficiently long times, including many episodes of rapid L-evolution).
   
   The particles trapped in the \ms\  of Pulsar B form the analogues of planetary radiation belts (van Allen belts). Both in planetary \mss\ and the Pulsar  B particles are injected at  the magnetosheath,  and then  diffuse inward.   The radial diffusion is  achieved by breaking of the 3d adiabatic invariant \citep{1968epf..conf..157F} and is described by radial diffusion equation
 \be
 \partial_t f = \partial_ L \left( D_{LL} {1\over L^2}  \partial_ L( L^2 f) \right) 
 \ee
 where $D_{LL} $ is the diffusion coefficient

 
The   diffusion coefficient $D_{LL} $ is a product of typical velocity times typical jump in $L$. 
The   diffusion is driven by the  E-cross-B velocity oscillations, approximately $\propto L$, Fig. \ref{1}. The induced diffusions is a non-adiabatic effect, proportional to the Larmor radius of the particles. 
Conservation of the first adiabatic invariant implies  betatron condition  $\epsilon_\perp \propto B$ - as a result, the Larmor radius remains nearly constant as a particle diffuse inward. 
 Thus,
 $D_{LL} = \kappa_0 L$, where $\kappa$ is some constant.
 Then the rate of change  of average value of $<L>$ is 
 \be
 \partial_t <L> \propto \kappa_0 =  {\rm constant}
 \ee
 while the  steady state is 
 \be
 f \propto \frac{1}{L}
 \ee
 Qualitatively, this is our main result: diffusion in fast corotating \mss\ produces density  of trapped particles  $\propto 1/r$.

\section{Simulations} 
\label{Simulations}

\subsection{Code verification} 

\begin{figure}[h!]
\includegraphics[width=.49\textwidth]{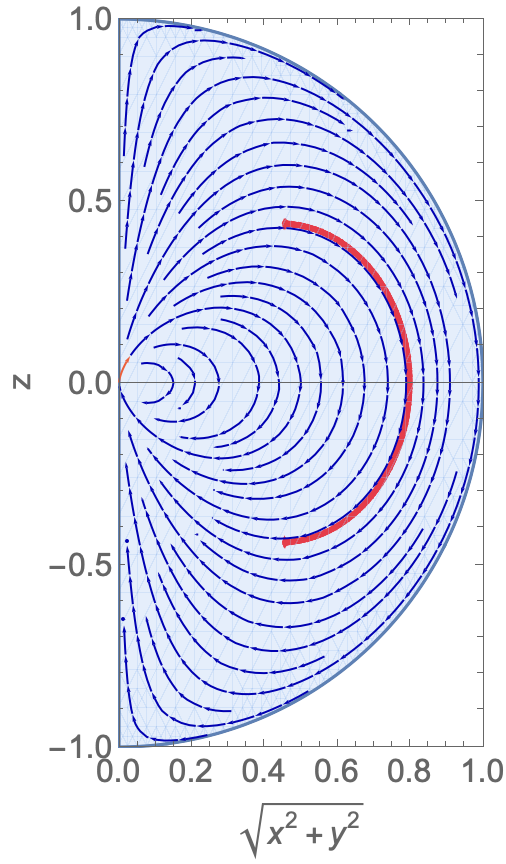}
\includegraphics[width=.49\textwidth]{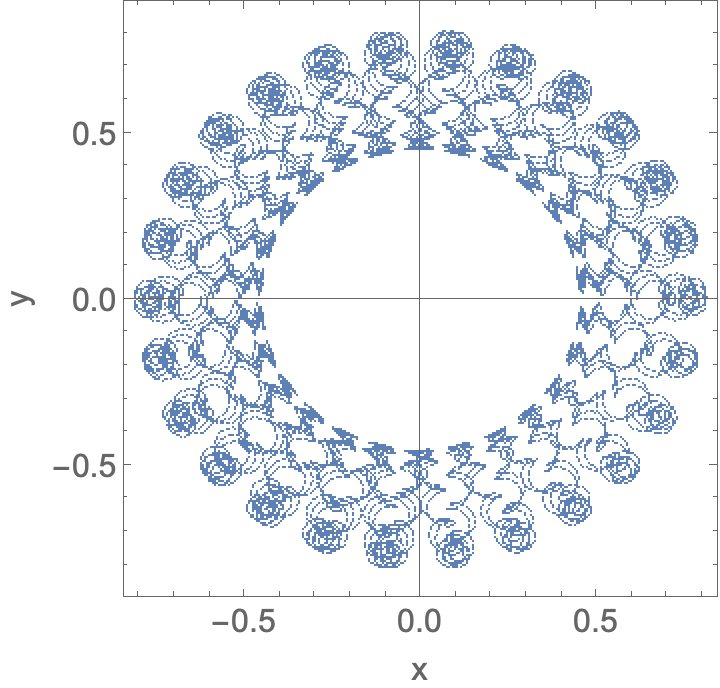}
\caption{Example of particle motion  calculated with our code in steady-state Dungey field ($z- \sqrt{x^2+y^2}$ and $x-y$ projections).  Particle is injected at $L=0.8$ at the equator. It experiences bouncing motion. The parameter $L$ remains constant }
\label{example} 
  \end {figure}

We have developed a Boris-based  pusher \citep{boris_69,birdsall}. Particles are injected at a given radius and move in the fields given by  (\ref{1}), 
 At each point the $L$ parameters is calculated according to (\ref{l}). We verified that for non-rotating  case $\Omega=0$  and sufficiently small Larmor radius parameter $L$ remains constant, Fig. \ref{example}.

We also verified that if initial velocity of particles is zero, they just move with the \EM\ drift velocity (\ref{1}). Non-zero initial momentum, even if directed along the local \Bf, leads to complicated particle dynamics with changing $L$ parameter. This is due to both effects of finite Larmor radius, and  various drifts.

   \subsection{Analysis of simulations}
     
\begin{figure}[h!]
\includegraphics[width=.49\textwidth]{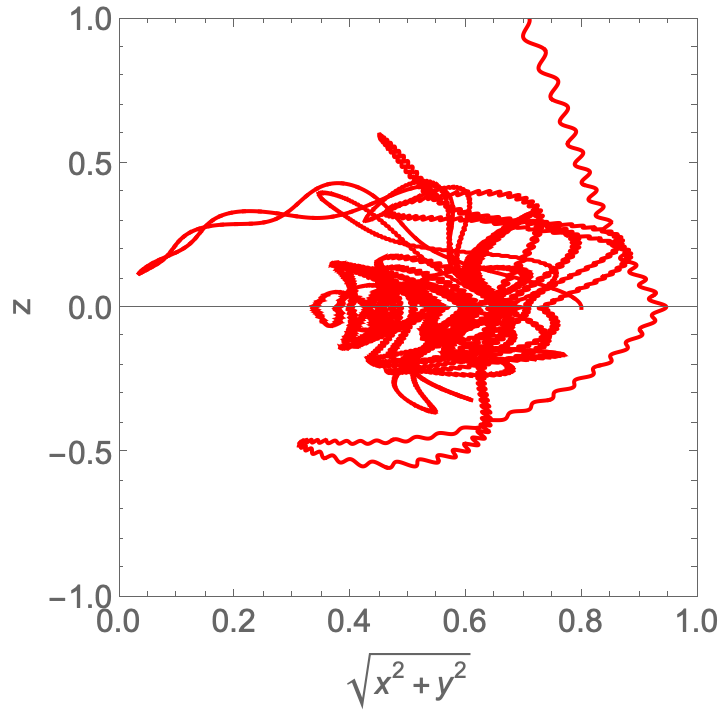}
\includegraphics[width=.49\textwidth]{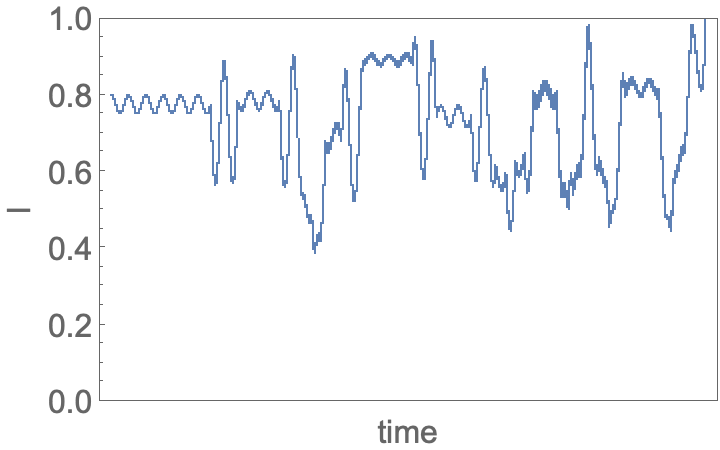}
\caption{An example demonstrating that  intervals  of just periodic oscillations are  often intermittent with  intervals of fast diffusion. This is expected, since the particle motion is controlled  by the Mathieu's equation (\protect\ref{Mathieu}) which has stable/unstable bands depending on the parameters.}
\label{example1} 
  \end {figure}

   For small gyration momenta  particles initially just follow the \Bf, with extra the drift motion. But eventually gets out of phase and the trajectory becomes random.
      We observe that the dynamics is not of a simple diffusion type: long intervals  of just periodic oscillations are interrupted by intervals of fast diffusion, Fig.  \ref{example1}.  We attribute these changes to the complicated structure of unstable bands of the   Mathieu's equation, see (\ref{Mathieu}).

We also observe that  some injected  particles often quickly escape. We attribute this to the fact that  the value of the E-cross-B drift is larger at larger radii. Thus, a particle oscillating along a given field line will have larger radial velocity when it is further out. 
  
\begin{figure}[h!]
\includegraphics[width=.49\textwidth]{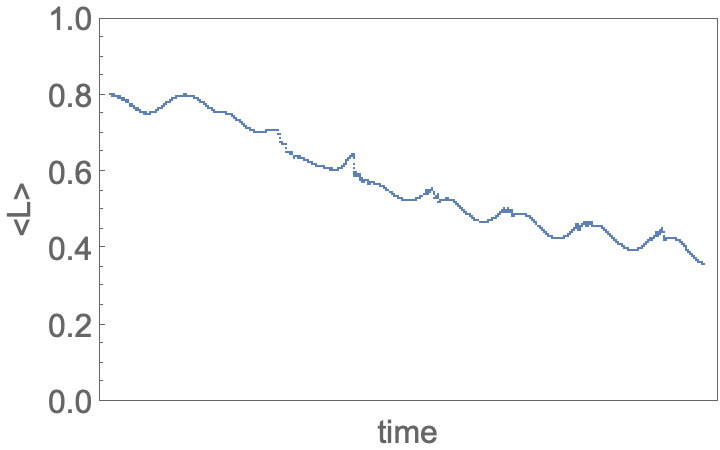}
\includegraphics[width=.49\textwidth]{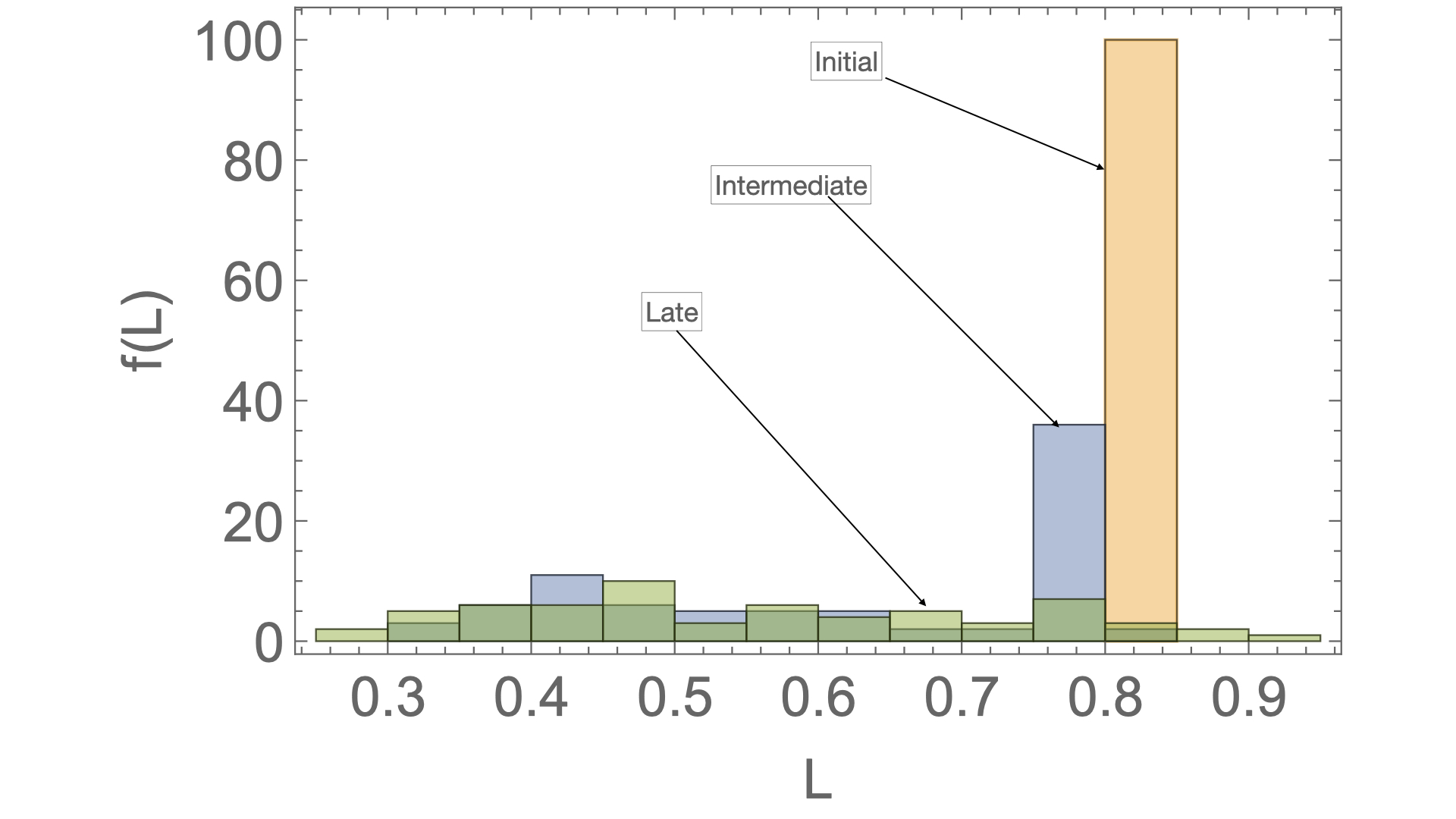}
\caption{Left Panel: Average value of the $L$ parameter as a function of time, showing that particles on average drift inwards.  Particles are injected at $L =0.8$ with the  same total momentum but  random direction. Right Panel:  evolution of the distribution $f(L)$ - particles are injected at $L=0.8$, and generally diffuse inward. }
\label{laveoft} 
  \end {figure}

One of the main points of the present work is illustrated in  Fig. \ref{laveoft}, which  shows that  particles drift inwards from the injection point.
  
  The diffusion is not homogeneous (in a sense that it does not obey the conventional diffusion equation), but stochastic: many particles remain at a fixed $L$, until they get into the regime of a parametric resonance.

 \section{Radial diffusion in rotating distorted \ms}
 \label{distorted3} 
 
Next  we complement our analysis with  a model of  distorted and rotating confined \mss, following  \cite{Stern94,1995JGR...100.5599T}, (previously  applied  to the Double Pulsar  by \cite{2005MNRAS.362.1078L}).  It is  a  different way  of modeling the distorted \ms: instead  of time-dependent underlying fields we construct a spatially-dependent  models of rotating confined magnetic configurations.  \citep[We also note a relevant paper by ][where a somewhat different model of the distorted planetary field was used. Importantly, the \ms\ was not co-rotating in that study.]{2003JGRA..108.1116E}

 First we   employ the method of distortion transformation of 
Euler potentials \citep{Stern94,1981P&SS...29....1V} to find \Bf. 
A major advantage of the stretching model of magnetosphere  is that it reproduces fairly 
well the structure of  a
tilted dipole \citep{Stern94}. Next we find rotationally induced \Ef, and then repeat out calculations of the trajectories. This method is somewhat different from the oscillating Dungey's \ms: in the case all the fields are stationary.

   \begin{figure}[h!]
\includegraphics[width=.49\textwidth]{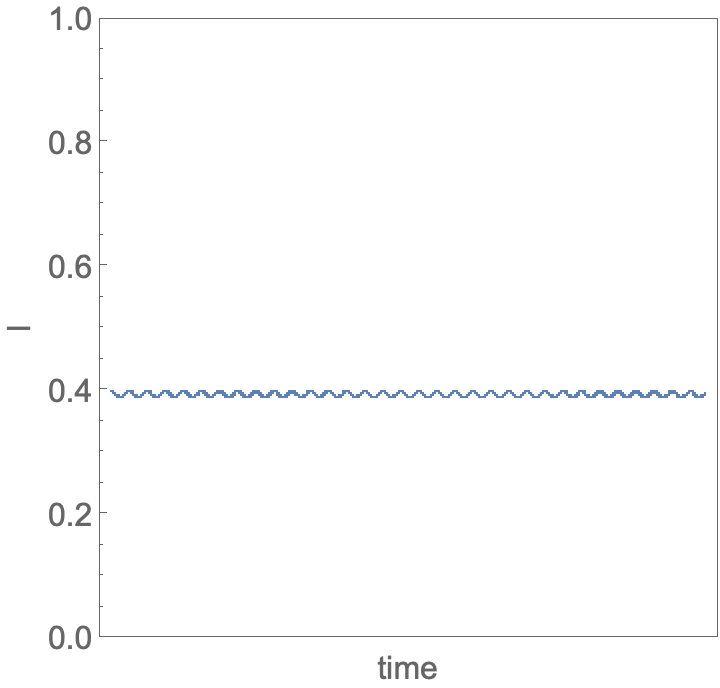}
\includegraphics[width=.49\textwidth]{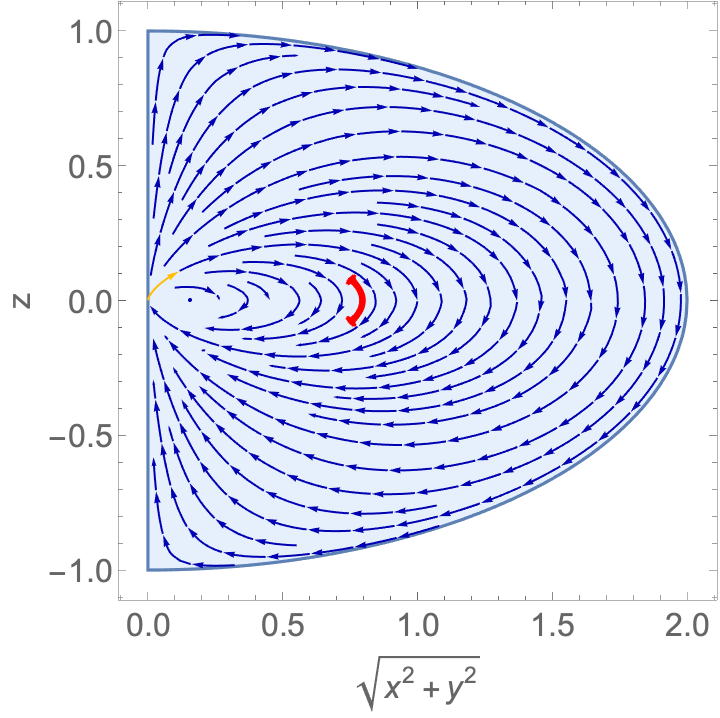}\\
\includegraphics[width=.49\textwidth]{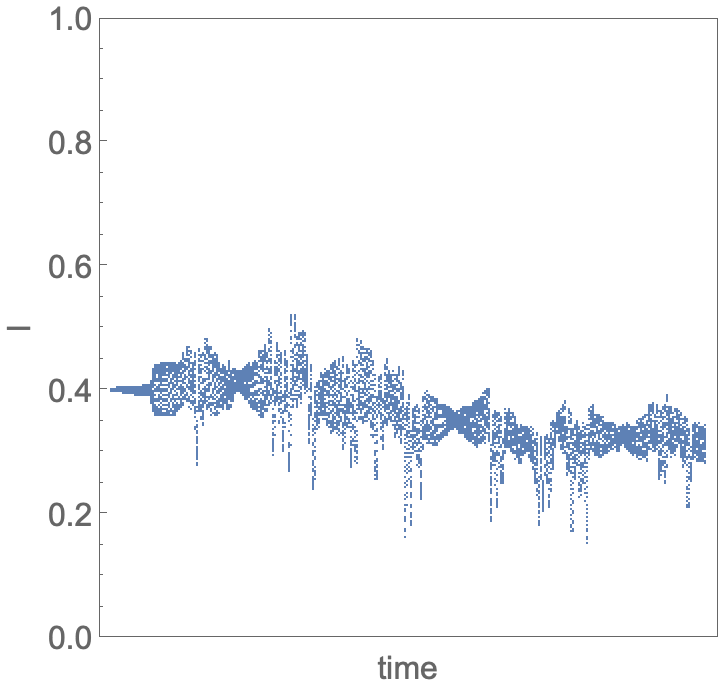}
\includegraphics[width=.49\textwidth]{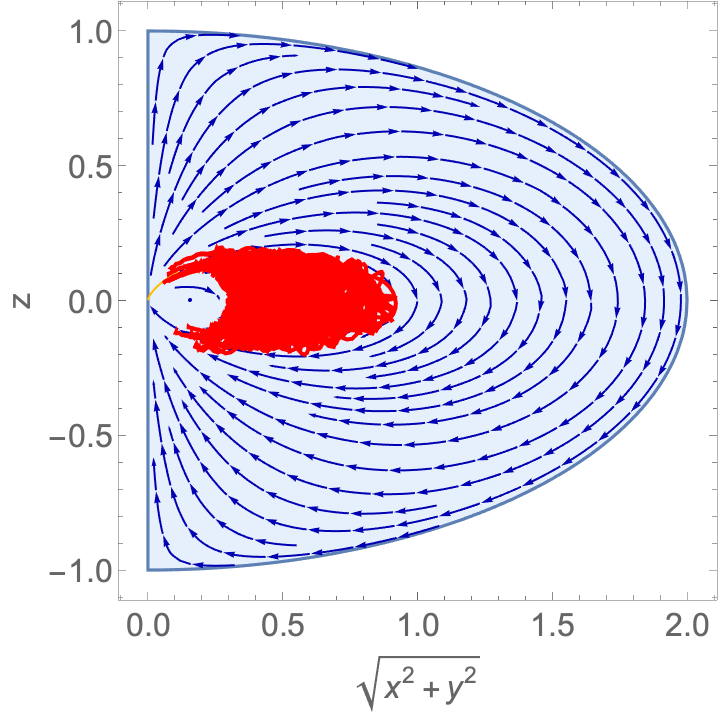}
\caption{Top row: $\Omega=0$ case.  Evolution of the L-parameter $L(t)$ (left panel)  and particle trajectory  $z(\sqrt{ x^2+y^2})$ (right panel). Bottom row: $\Omega =0.05$.  Same initial velocities, distortion parameter $C=1/2$. Initial location corresponds to $L=0.4$.}
\label{distorted} 
  \end {figure}

Magnetic field can  be  described by two 
Euler potentials $\alpha$ and $\beta$ (sometimes  called Clebsh potentials):
\be
{\bf B} = \nabla \alpha \times  \nabla \beta
\ee
so that magnetic field line is defined by an
intersection of  surfaces with constant 
 $\alpha$ and $\beta$.
Magnetosphere  of B enshrouded by   magnetopause resembles
a dipole field compressed on the dayside and stretched out on the nightside. The structure
of the nightside 
magnetosphere 
can be approximated  by stretching transformations of the 
Euler potentials $\alpha$ and $\beta$.

In axially-symmetric \ms,
\ba &&
\alpha = \Phi,\,   \mbox{Eq. (\protect\ref{1}), with constant $R_0$}
\nn && 
\beta = \phi
\ea
Converting to Cartesian coordinates, stretching the $x$ coordinate by transformation $ x \to  C x$ we find the new Euler potentials and the \Bf\  $\B_1$
(relations become cumbersome so we omit then here: the procedure is clear).  By construction the new field $\B_1$  satisfies $\div \B_1=0$. 
The shape of the cavity is 
\be
C^2 x^2 +y^2 +z^2 =R_0^2
\ee
On the surface the normal component of the \Bf\ vanishes.

 The $L$ parameter (\ref{l}) is now defined as 
 \be 
\frac{\left(x^2+y^2\right) \left(R_0^3-\left(C^2
   x^2+y^2+z^2\right)^{3/2}\right)}{\left(x^2+y^2+z^2\right) \sqrt{C^2
   x^2+y^2+z^2}}=\frac{1-L^3}{L}
   \ee

As  novel step, let's assume  that the central star is rotating. Purely for simplicity, and to demonstrate the principal effect, let's assume that the star's spin is along $z$ axis, so that inside the cavity the plasma moves with constant coordinate $z$. 
This requires 
\be
C^2 x \beta_x + y \beta_y =0
\ee
Thus
\ba &&
\beta_x = -\frac{y \Omega  \sqrt{x^2+y^2}}{C^2 x \sqrt{\frac{y^2}{C^4
   x^2}+1}}
   \nn &&
   \beta_y= \frac{\Omega  \sqrt{x^2+y^2}}{\sqrt{\frac{y^2}{C^4 x^2}+1}}
   \ea
   
   The \Ef\ is then
   \be
   \E_1 = - {\bf  \beta}  \times \B_1
   \ee
Using fields $B_1$ and $E_1$ in the Boris pusher, we find dynamics very similar to the case of oscillating Dungey \ms, Fig. \ref{distorted}.

  The  results of the two approaches (oscillating Dungey \ms, and the stretched out rotating \ms)    show qualitatively the same picture: strong radial diffusion is induced by the rotating \ms.  
    Thus, using two complementary methods: oscillating Dungey's \ms\ and rotating stationary \ms\ we arrive at the same result: betatron-type induced radial diffusion.
 
 \section{Application to   Pulsar B}

Let us assume that injection rate is fraction of influx of the Pulsar A wind hitting the Pulsar B \ms.  Pulsar A's spindown power can be related to the particle influx at  Pulsar B \ms:
\be
L_A = 4\pi r_{AB}^2 n \Gamma_w m c^3
\ee
where $L_A$ is Pulsar A's spindown power,  $ r_{AB}$ is the distance between two pulsars, and $\Gamma_w$ is  the  \Lf\ of the Pulsar A's wind ($\sigma \sim 1$ is assumed for estimates).

The \ms\ of Pulsar B covers approximately $10^{-3}$ of $ 4 \pi$, as seen from Pulsar A  \citep{lt05}. Let's assume that  efficiency of getting into \ms\ is $\zeta$ (it likely depends on the orbital phase, \eg\ as a fraction of polar cap ``facing the wind''). The injection rate is then
\be 
\dot{N} = {L_A \eta \zeta \over \Gamma_w m c^2}
\label{dotN} 
\ee

The density of particles in the \ms\ is determined by the balance of the  injection rate (\ref{dotN}) and the rate at which particles fall onto the star due to radiative losses. 
Radiative decay rate is  $N/(\kappa P_B)$ 
\ba &&
\dot{N} =  N/(\kappa P_B)
\nn &&
N= \dot{N} \kappa P_B
\ea
where 
\be
\kappa = \frac{1}{8} \left(  \frac{ R_m}{ R_{cool}}  \right)^{-3}  \sim 2\times 10^{-5}
\ee
is the  inverse of  a typical time (in periods of  Pulsar B) that particles spend in the \mss, $R_m = 4 \times 10^9$ cm  is the radius of the wind-confined \ms\ of Pulsar B, $R_{cool}= 2.4 \times 10^{8} $ cm is the cooling radius (where cooling time becomes of  the order of the Pulsar  B period), \cite{lt05}.

The equilibrium density is then 
\ba &&
n = { N \over ( 4\pi /3) R_m^3}= { \dot{N} \kappa P_B \over  ( 4\pi /3) R_m^3}
\nn &&
\lambda=  { n \over n_{GJ}} = {3  \pi e L_A \eta \kappa \zeta \over B_m c m R_m ^3 \Gamma_w \Omega_B  ^2}
 ={ 3\times 10^8 \over \Gamma_w/(10^5)}
 \label{lambda} 
 \ea
 ($\lambda $ is the over-density with respect to the \cite{GJ})  density $n_{GJ}$). 
 The estimate   (\ref{lambda})  exceeds the  minimal  density required imposed by the eclipse model of  \cite{lt05}. 

 \section{Discussion}

In this paper we discuss an unusual regime of radial diffusion in the  wind-confined \mss: when the secondary (Pulsar B in this case) is fast rotating, so that the rotational period is shorter than the time scale of  azimuthal drifts. In this regime  the third adiabatic invariant is strongly violated just by the \EM\ fields arising from rotational compression of the \ms.  No extra turbulence is needed. We demonstrate that in this case the radial diffusions is driven purely by the rotating of the central star - there is  no need for turbulent pitch angle scattering to induce the radial diffusion. The  radial diffusion occurs to what can be  called  ``a betatron-induced diffusion": fluctuations of the \Ef\ (either in Lagrangian or Eulerian sense - the two models we considered) induce  (occasional) parametric instability in particle's orbits.

Thus,  the  magnetospheric  boundary is  indeed where \cite{lt05} calculated it to be. But the density of the trapped practices increases inward - hence smaller eclipsing region,  by a factor of $\sim 3$.

\section{ACKNOWLEDGEMENTS}

I would like to thank Mary  Hudson for comments on the manuscript and Maura McLaughlin for discussions.
This work had been supported by NASA grants 80NSSC17K0757 and 80NSSC20K0910, NSF grants 1903332 and 1908590.

\section{DATA AVAILABILITY}
The data underlying this article will be shared on reasonable request to the corresponding author.

\bibliographystyle{apj}
  \bibliography{/Users/maxim/Home/Research/BibTex}

 \end{document}